\documentclass[doublecol]{epl2} 
\usepackage{color}

\title{Effects of Lorentz Symmetry Violation on a Relativistic Scalar Particle in Quantum Systems}
\shorttitle{Scalar Particle under LSV} 

\author{FAIZUDDIN AHMED\thanks{E-mail: \email{faizuddinahmed15@gmail.com}}}
\shortauthor{F. Ahmed \etal}

\institute{                    
          National Academy Gauripur, Assam, 783331, India
}

\pacs{11.30.Cp}{Lorentz and Poincar\'{e} invariance}
\pacs{03.65.Pm}{Relativistic wave equations}
\pacs{11.30.Qc}{Spontaneous and radiative symmetry breaking}

\abstract{
In this paper, the relativistic quantum dynamics of a scalar particle under the effect of Lorentz symmetry violation determined by a tensor $(K_F)_{\mu\nu\alpha\beta}$ out of the Standard Model Extension is investigated. We see that the bound-state solutions of the modified Klein-Gordon equation can be obtained, and the spectrum of energy and the wave function depends on the Lorentz symmetry breaking parameters.}

\begin{document}

\maketitle

\section{Introduction}

The relativistic quantum dynamics of scalar particles, like bosons by solving the Klein-Gordon equation with (-out) potential of different types in various spacetime background have been investigated in quantum systems (see, Refs. \cite{b.a1,b.a2,b.a3,b.a4,b.a5,b.a6,b.a7,b.a8} and related references therein). In the present work, we investigate the quantum motion of a scalar particle under the effect of Lorentz symmetry breaking defined by a tensor $(K_F)_{\mu\nu\alpha\beta}$ out of the Standard Model Extension (SME). We analyze the effects on the energy eigenvalue and the wave function and see that the result gets modified in comparison to the standard Landau levels. We also see that the bound-state solutions of the modified Klein-Gordon equation can be obtained. 

In quantum systems, effect of Lorentz symmetry violation (LSV) have been investigated, for example, in the nonrelativistic limit, the spectrum of the hydrogen atom \cite{b.a9}, quantum Hall conductivity Ref. \cite{b.a10}, the Aharonov-Bohm-Casher effect Ref. \cite{b.a11}, neutral Dirac particles Refs. \cite{b.a12,b.a13,b.a14,b.a15}, Rashba-like couplings Refs. \cite{b.a14,b.a16}, Landau-type system Ref. \cite{b.a17}, geometric quantum phases Refs. \cite{b.a12,b.a13,b.a18,b.a19}, and the harmonic oscillator Ref. \cite{b.a20} ; in the relativistic limit, scalar particles Ref. \cite{b.a21,b.a22, b.a23,b.a24}, the quantum oscillator Ref. \cite{b.a25}, Landau–He–McKellar–Wilkens quantization and Dirac particles Ref. \cite{b.a26}, the Klein–Gordon oscillator Ref. \cite{b.a27}, the Dirac oscillator Ref. \cite{b.a28}, the relativistic geometric quantum phases Refs. \cite{b.a29, b.a30,b.a31}, scalar particles subject to a scalar potential in cosmic string space-time Ref. \cite{b.a32}, relativistic EPR (Einstein–Podolsky–Rosen) correlations Ref. \cite{b.a33}, a massive scalar field under the influence of a Coulomb-type and linear central potential Ref. \cite{b.a34}, and a modified Klein-Gordon oscillator under a linear central potential Ref. \cite{b.a35}.

A possible way of dealing with a scenario beyond the Standard Model is the extension of the mechanism for the spontaneous symmetry breaking through a tensor fields, which implies the violation of the Lorentz symmetry Ref. \cite{b.a36}. It is shown that the Lorentz symmetry is violated through a spontaneous symmetry breaking mechanism triggered by the appearance of nonvanishing vacuum expectation values of tensor fields. A general framework for testing the low-energy manifestations of the CPT-symmetry and the Lorentz symmetry breaking is known as the Standard Model Extension (SME) Refs. \cite{b.a37,b.a38,b.a39,b.a40}, an extension of the Standard Model of fundamental interactions. In this framework, tensor-valued background fields are suitably contracted with the Standard Model field operators and these observer Lorentz scalars are added to the Lagrangian density of the Standard Model. The current limits on the coefficients of the Lorentz symmetry violation can be found in detail in Ref. \cite{b.a41}.

There are two types or classes of Lorentz-violating contributions in the electromagnetic sector of the SME that modify the propagation properties of electromagnetic waves in spacetime. These two classes are called the CPT-odd sector Refs. \cite{b.a37,b.a38} and the CPT-even sector Refs. \cite{b.a42,b.a43, b.a44,b.a45,b.a46,b.a47,b.a48}. The Lorentz symmetry violation established by a tensor field is introduced into the Klein–Gordon equation through a nonminimal coupling given by $\frac{g}{4}\,(K_F)_{\mu\nu\alpha\beta}\,F^{\mu\nu}(x) \,F^{\alpha\beta}(x)$, where $g$ is the coupling constant whose mass dimension is $-2$, $F_{\mu\nu} (x)$ is the electromagnetic (EM) field tensor, and $(K_F)_{\mu\nu\alpha\beta}$ corresponds to the CPT-even tensor of the electromagnetic sector of the SME Refs. \cite{b.a37,b.a38,b.a39,b.a40} and it is a dimensionless tensor. The tensor $(K_F)_{\mu\nu\alpha\beta}$ has the symmetries of the Riemann tensor $R_{\mu\nu\alpha\beta}$ and zero on double trace $(K_F)^{\mu\nu}_{\,\mu\nu}=0$, so it contains 19 independent real components (see, Refs. \cite{b.a37,b.a38, b.a44,b.a45,b.a46,b.a47,b.a48} for details). 

Therefore, the relativistic quantum dynamics of a scalar particle under the effects of LSV Refs. \cite{b.a21,b.a22, b.a23,b.a24,b.a25,b.a27,b.a29,b.a37,b.a38,b.a45,b.a46,b.a47} is given by
\begin{equation}
\left [-p^{\mu}\,p_{\mu}+\frac{g}{4}\,(K_F)_{\mu\nu\alpha\beta}\,F^{\mu\nu} (x)\,F^{\alpha\beta}(x) \right]\,\Psi= M^2\, \Psi,
\label{1}
\end{equation}
where $\Psi$ is the scalar wave function and $M$ is the rest mass of the particle \footnote{In the sign convention $(-,+,+,+)$, momentum four-vector is defined $p^{\mu}=(E, \vec{p})$. So, its covariant form is given by $
p_{\mu}=(-E, \vec{p})$. Thus, $-p^{\mu}\,p_{\mu}=-(-E^2+\vec{p}^{\,2})=M^2$ using the mass-energy relation. Here, we have chosen the units $c=1=\hbar$}.

The structure of this paper is as follows: in {\it section 2}, we establish the background of Lorentz symmetry violation defined by the tensor $(K_F)_{\mu\nu\alpha\beta}$ out of the SME. Then, we analyze the behaviour of a relativistic scalar particle by solving the modified Klein-Gordon equation; in {\it section 3}, we present our conclusions.

\section{Lorentz Symmetry Breaking Effects on a Relativistic Scalar Particle}

We consider Minkowski spacetime in cylindrical coordinates $(t, r, \phi, z)$
\begin{equation}
ds^2=-dt^2+dr^2+r^2\,d\phi^2+dz^2,
\label{2}
\end{equation}
where the ranges of the coordinates are $-\infty < (t, z) < \infty$, $r\geq 0$ and $ 0 \leq \phi \leq 2\,\pi$. 

For the geometry (\ref{2}), the modified Klein-Gordon equation (\ref{1}) becomes
\begin{eqnarray}
&&\left[-\frac{\partial^2}{\partial t^2}+\frac{\partial^2}{\partial r^2}+\frac{1}{r}\,\frac{\partial}{\partial r}+\frac{\partial^2}{\partial z^2}+\frac{1}{r^2}\,\frac{\partial^2}{\partial \phi^2}\right]\,\Psi\nonumber\\
&+&\frac{g}{4}\,(K_F)_{\mu\nu\alpha\beta}\,F^{\mu\nu} (x)\,F^{\alpha\beta}(x)\,\Psi=M^2\,\Psi.
\label{3}
\end{eqnarray}

Based on Refs. \cite{b.a44,b.a45,b.a46,b.a47,b.a48}, the tensor $(K_F)_{\mu\nu\alpha\beta}$ can be written in terms of $3 \times 3$ matrices that represent the parity-even sector: $(\kappa_{DE})_{jk}=-2\,(K_F)_{0j0k}$ and $(\kappa_{HB})_{jk}=\frac{1}{2}\,\epsilon_{jpq}\,\epsilon_{klm}\,(K_F)^{pqlm}$, and the parity-odd sector: $(\kappa_{DB})_{jk}=-(\kappa_{HE})_{kj}=\epsilon_{kpq}\,(K_F)^{0jpq}$. The matrices $(\kappa_{DE})_{jk}$ and $(\kappa_{HB})_{jk}$ are symmetric and the matrices $(\kappa_{DB})_{jk}$ and $(\kappa_{HE})_{kj}$ have no symmetry. In this way, we can rewrite (\ref{3}) in the form :
\begin{eqnarray}
&&\left[-\frac{\partial^2}{\partial t^2}+\frac{\partial^2}{\partial r^2}+\frac{1}{r}\,\frac{\partial}{\partial r}+\frac{1}{r^2}\,\frac{\partial^2}{\partial \phi^2}+\frac{\partial^2}{\partial z^2}\right]\,\Psi\nonumber\\
&+&\left[-\frac{g}{2}\,(\kappa_{DE})_{ij}\,E^{i}\,E^{j}+\frac{g}{2}\,(\kappa_{HB})_{ij}\,B^i\,B^j \right]\,\Psi\nonumber\\
&&-g\,(\kappa_{DB})_{ij}\,E^i\,B^j\,\Psi=M^2\,\Psi.
\label{4}
\end{eqnarray}

Let us consider a possible scenario of Lorentz symmetry violation determined by $(\kappa_{DE})_{r\,r}=const=\kappa_1$, $(\kappa_{HB})_{z\,z}=const=\kappa_2$, $(\kappa_{DB})_{r\,z}=const=\kappa_3$ with the configuration of electric and magnetic fields given by Refs. \cite{b.a13,b.a24,b.a27,b.a28}:
\begin{equation}
\vec{B}=B\,\hat{z}\quad,\quad \vec{E}=\frac{\lambda}{r}\,\hat{r},
\label{5}
\end{equation}
where $B>0$, $\hat{z}$ is a unit vector in the $z$-direction, $\lambda$ is the linear charge density of the electric charge distribution, and $\hat{r}$ is the unit vector in the radial direction. 

Hence, equation (\ref{4}) using the configuration (\ref{5}) becomes
\begin{eqnarray}
&&\left[-\frac{\partial^2}{\partial t^2}+\frac{\partial^2}{\partial r^2}+\frac{1}{r}\,\frac{\partial}{\partial r}+\frac{1}{r^2}\,\frac{\partial^2}{\partial \phi^2}+\frac{\partial^2}{\partial z^2} -\frac{g\,\kappa_1\,\lambda^2}{2\,r^2}\right]\,\Psi\nonumber\\
&&+\left[\frac{g\,\kappa_2\,B^2}{2}-\frac{g\,\kappa_3\,\lambda\,B}{r}-M^2\right]\,\Psi=0.
\label{6}
\end{eqnarray}

Since the metric (\ref{2}) is independent of time and symmetric with respect to translations along the $z$-axis, as well as with respect to rotations, it is reasonable to write the total wave function $\Psi (t, r, \phi, z)$ in terms of the radial wave function $\psi (r)$ as follows:
\begin{equation}
\Psi (t, r, \phi, z)=e^{i\,(-\varepsilon\,t+m\,\phi+k\,z)}\,\psi (r),
\label{7}
\end{equation}
where $\varepsilon$ is the energy of the scalar particle, $m=0,\pm 1,\pm 2,....$ are the eigenvalues of the $z$-component of the angular momentum operator, and $k$ is a constant.

Substituting the solution (\ref{7}) into the Eq. (\ref{6}), we obtain the following radial wave equation for $\psi(r)$:
\begin{equation}
\psi''(r)+\frac{1}{r}\,\psi' (r)+\left[-\Lambda^2-\frac{j^2}{r^2}-\frac{\delta}{r} \right]\,\psi (r)=0,
\label{8}
\end{equation}
where
\begin{eqnarray}
&&\Lambda^2=M^2+k^2-\varepsilon^2-\frac{1}{2}\,g\,B^2\,\kappa_2,\nonumber\\
&&j=\sqrt{m^2+\frac{1}{2}\,g\,\lambda^2\,\kappa_1},\nonumber\\
&&\delta=g\,\lambda\,B\,\kappa_3.
\label{9}
\end{eqnarray}

Let us now perform a change of variables via $x=2\,\Lambda\,r$. Then, from Eq. (\ref{8}) we have 
\begin{equation}
\psi''(x)+\frac{1}{x}\,\psi' (x)+\left(-\frac{1}{4}-\frac{j^2}{x^2}-\frac{\delta}{2\,\Lambda\,x} \right)\,\psi (x)=0,
\label{10}
\end{equation}

By analysing the asymptotic behaviour of Eq. (\ref{10}) for $x \rightarrow 0$ and $x \rightarrow \infty$, we can write a possible solution to the Eq. (\ref{10}) that can be expressed in terms of an unknown function $F (x)$ as
\begin{equation}
\psi (x)=x^{j}\,e^{-\frac{x}{2}}\,F (x).
\label{11}
\end{equation}
Thereby, substituting the solution (\ref{11}) into the Eq. (\ref{10}), we obtain the following differential equation :
\begin{equation}
x\,F'' (x)+(1+2\,j-x)\,F' (x)+\left(-\frac{\delta}{2\,\Lambda}-j-\frac{1}{2} \right)\,F (x)=0,
\label{12}
\end{equation}
which is called the confluent hyper-geometric equation Refs. \cite{b.a49,b.a50}. The function $F(x)$ is the confluent hyper-geometric function, that is, $F(x) = \,_{1}F_{1} (j+\frac{1}{2}+\frac{\delta}{2\,\Lambda}, 2\,j+1, x)$. It is well-known that the confluent hyper-geometric series becomes a polynomial of degree $n$ when $\left(j+ \frac{1}{2}+\frac{\delta}{2\,\Lambda}\right)=-n$ Refs. \cite{b.a49,b.a50} where, $n=0,1,2,3,....$. 

After simplification of $\left (j+\frac{1}{2}+\frac{\delta}{2\,\Lambda} \right)=-n$, we have obtained the energy eigenvalue $\varepsilon_{n,m}$ as follows: 
\begin{equation}
\varepsilon_{n,m}=\pm\,\sqrt{M^2+k^2-\frac{1}{2}\,g\,B^2\,\Delta},
\label{13}
\end{equation}
where $\Delta=\left( \kappa_2+\frac{g\,\lambda^2\,\kappa^{2}_3}{2\,\left(n+\frac{1}{2}+\sqrt{m^2+\frac{1}{2}\,g\,\lambda^2\,\kappa_1}\right)^2} \right)$.

The radial wave function is given by
\begin{equation}
\psi_{n,m} (x)=x^{\sqrt{m^2+\frac{1}{2}\,g\,\lambda^2\,\kappa_1}}\,e^{-\frac{x}{2}}\,\,_{1}F_{1} (j+\frac{1}{2}+\frac{\delta}{2\,\Lambda}, 2\,j+1, x).
\label{14}
\end{equation}

Equation (\ref{13}) gives the allowed values of the energy eigenvalue of a relativistic scalar particle under the effects of LSV. We have established the background of LSV by a tensor possessing the nonzero components $(\kappa_1, \kappa_2, \kappa_3)$, a radial electric field produced by a linear distribution of the electric charges, and a constant magnetic field along the $z$-direction. One can easily show for $\kappa_1=0$ and $\kappa_2=0$ that the energy eigenvalue of Eq. (\ref{13}) and the wave function of Eq. (\ref{14}) reduces to the result obtained in Ref. \cite{b.a24}. Thus, the result presented in this work is modified in comparison to the results found in Ref. \cite{b.a24}. The negative energy of Eq. (\ref{13}) indicates that the energy of a spin-$0$ scalar particle is symmetrical about $\varepsilon_{n,m}=0$ for constant or zero values of $m$. However, in general the negative energy of scalar particles-like bosons can not be explained by the Klein-Gordon or the modified Klein-Gordon theory. One might deal with the Dirac equation (applicable for half-spin particles only) and that requires a multi-particle theory (which quantum mechanics is not).

\section{Conclusions}

We have investigated the effects of Lorentz symmetry violation on a relativistic scalar particle by solving the Klein-Gordon equation in the context of relativistic quantum systems. Though our work is inspired by SME that is an effective field theory, in our analysis we have relaxed the renormalization property. We have shown that the  bound-state solutions of the Klein-Gordon equation can be obtained for a radial electric field produced by a linear distribution of the electric charge, a uniform magnetic field along the $z$-direction, and the dimensionless  Lorentz-violating tensor $(K_F)_{\alpha\beta\mu\nu}$ possessing the nonzero components $(\kappa_1, \kappa_2, \kappa_3)$. It is worth mentioning that the above nonzero components of the Lorentz symmetry breaking tensor are constant in the coordinate system determined by the line element (\ref{2}) {\it i. e.} in cylindrical coordinates $(t, r, \phi, z)$. There is nothing that forbids this assumption. However, if one changes the coordinate system, then the components of the Lorentz symmetry breaking terms are no longer constant breaking momentum conservation.  Studying the effects related to momentum nonconservation is beyond the scope of the present article. After solving the radial wave equation, we have obtained the energy eigenvalues Eq. (\ref{13}) and the wave function Eq. (\ref{14}) of a relativistic scalar particle. We can see that the presence of the Lorentz symmetry breaking parameters $(g, B, \lambda, \kappa_1, \kappa_2, \kappa_3)$ implies the modified energy spectrum by Eq. (\ref{13}) and the wave function by Eq. (\ref{14}) in comparison to the results obtained in Ref. \cite{b.a24}.

\acknowledgements

I sincerely acknowledge the kind referee(s) for his/her valuable comments and suggestions.

\end{document}